\newtheorem{theorem}{Theorem}
\newtheorem{proposition}[theorem]{Proposition}

\newtheorem{cor}[theorem]{Corollary}

\newtheorem{remark}[theorem]{Remark}
\newcommand{\pf}{{\noindent \bf Proof. \ }}
\newcommand{\qed}{\hfill $\Box$ \\}
\newcommand{\C}{C}

\newcommand{\R}{{ R}}

\newcommand{\x}{{\bf x}}
\newcommand{\y}{{\bf y}}
\newcommand{\z}{{\bf z}}
\newcommand{\cc}{{\bf c}}

\documentclass[journal]{IEEEtran}
\usepackage{amsmath,amssymb,amsfonts}

\begin{document}
\title{Classification of extremal and $s$-extremal binary self-dual codes of length $38$}

\author{Carlos Aguilar-Melchor,  Philippe Gaborit, Jon-Lark Kim, Lin Sok and Patrick Sol\'{e}%
\thanks{C. Aguilar-Melchor and P. Gaborit are with XLIM-DMI, UMR 6172, Universit\'e de Limoges,
123, Av. Albert Thomas,
 87000 Limoges, France. {\tt carlos.aguilar@xlim.fr,gaborit@unilim.fr}}

\thanks{J.-L. Kim, Department of Mathematics, University of Louisville, Louisville, KY 40292, USA. {\tt jl.kim@louisville.edu}}

\thanks {Lin Sok, Department of Comelec, Telecom ParisTech, 46 rue Barrault 75013 Paris, France.
{\tt lin.sok@telecom-paristech.fr}}

\thanks {Patrick Sol\'{e}, CNRS/LTCI, UMR 5141, Telecom ParisTech 46
rue Barrault 75 634 Paris cedex 13, France and MECAA, Math Dept of
King Abdulaziz University, Jeddah, Saudi Arabia {\tt
patrick.sole@telecom-paristech.fr}} }

%\author{\authorblockN{Carlos Aguilar Melchor}
%\authorblockA{ XLIM-DMI, Universit\'e de Limoges,\\
%123 av. Albert Thomas, \\
%87000, Limoges, France \\
%Email: carlos.aguilar@unilim.fr}
%\and
%\authorblockN{Philippe Gaborit}
%\authorblockA{
%XLIM-DMI, Universit\'e de Limoges,\\
%123 av. Albert Thomas, \\
%87000, Limoges, France \\
%Email: gaborit@unilim.fr}}\bibitem{XQ48} S. Houghten,  C. Lam, L. Thiel and J. Parker,

\maketitle

\begin{abstract}

In this paper we classify all extremal and $s$-extremal binary self-dual codes of length $38$.
There are exactly $2744$ extremal $[38,19,8]$ self-dual
codes, two $s$-extremal $[38,19,6]$ codes, and
$1730$ $s$-extremal $[38,19,8]$ codes. We obtain our results from the
use of a recursive algorithm used in the recent classification of
all extremal self-dual codes of length 36, and from a generalization of this
recursive algorithm for the shadow.
The classification of $s$-extremal $[38,19,6]$ codes permits
to achieve the classification of all $s$-extremal codes with $d=6$.

\end{abstract}

\medskip


%\maketitle

\section{Introduction}

Self-dual codes are one of the most interesting classes of linear codes. They have close connections with group theory, lattice theory, design theory, and modular forms. It is well known that self-dual codes are asymptotically good~\cite{MacSloTho}. There has been an
active research on the classification of self-dual codes over finite fields and over rings in general (see~\cite{RS},~\cite{NRS} for details). In particular, the classification of binary self-dual codes was started by Pless~\cite{Ple-F2} and has been actively studied by many authors (see~\cite{Huffman} for a survey of optimal self-dual codes over small alphabets).

Recently, using a recursive method, Aguilar and Gaborit classified
all $41$ extremal $[36,18,8]$ binary self-dual codes. These results were pushed
further by Harada and Munemasa~\cite{HarMun_10} who, besides the 41 extremal codes of
\cite{AG}, also give a complete classification of all self-dual codes of length $36$.

A natural question is hence to consider the case of length $38$.
A simple computation on the mass formula shows that there are at least $13,644,433$ inequivalent
binary self-dual $[38, 19]$ codes~\cite{HarMun_10}.
It is hence natural to consider the case of special subclasses of self-dual codes.
The most interesting such subclass is the class of extremal codes.
Given the classification of all $[36,18,6]$ self-dual
codes of \cite{HarMun_10}, we apply an optimized recursive algorithm as in \cite{AG}
to derive the classification of all $2744$ extremal self-dual $[38,19,8]$ codes.

Another subclass of interesting self-dual codes with combinatorial properties is the class of $s$-extremal codes: these codes are self-dual
codes whose weight enumerator is uniquely determined, depending on the condition on a high weight of the shadow. The notion of codes (and lattices) with long shadows was first developed by Elkies \cite{E95}. This notion was generalized by Bachoc and Gaborit in \cite{BG} who introduced the notion of $s$-extremal codes. These codes exist depending on conditions on their length
and their minimum distance. The classification of $s$-extremal codes with $d=4$ was done
by Elkies. The case of $d=6$ was mainly considered in \cite{BG}, but two lengths
remained to be classified. One is length $36$, which was classified in \cite{AG}, and the other is
length $38$, which is what we classify in this paper.
Our classification is based on a generalization of the subtraction algorithm
in the case of the shadow. It permits us to use the recursive algorithm by showing
that in certain cases for $n$ even, the subtraction of $(11)$ from
a $[2n+2,n+1,d+2]$ self-dual code with shadow weight $s+1$ leads to
a $[2n,n,d]$ self-dual code with shadow weight $s$. This result is interesting in itself.

The paper is organized as follows: Section II gives preliminaries and background for self-dual codes, Section III compares the different method to extend a self-dual code in a purpose of classification. In Section IV we show that there are exactly $2744$ extremal $[38,19,8]$ binary self-dual codes.
In Section V we prove that there are only two $s$-extremal $[38, 19, 6]$ codes and $1730$ $s$-extremal $[38, 19, 8]$ codes. The last section describes the covering radii of self-dual codes of length $38$.

\section{Preliminaries}
\label{Sec-Prelim}

We refer to~\cite{HufPle} for basic definitions and results related
to self-dual codes. All codes in this paper are binary. A {\em
linear $[n,k]$ code $C$ of length $n$} is a $k$-dimensional subspace
of  $ GF(2)^n$. An element of $C$ is called a {\em codeword}. The
(Hamming) weight wt$({\bf{x}})$ of a vector ${\bf{x}}=(x_1, \dots,
x_n)$ is the number of non-zero coordinates in it. The {\em minimum
distance ({\rm{or}} minimum weight) $d(C)$} of $C$ is
$d(C):=\min\{{\mbox{wt}}({\bf{x}})~|~ {\bf{x}} \in C, {\bf{x}} \ne
{\bf{0}} \}$. The {\em Euclidean inner product} of ${\bf{x}}=(x_1,
\dots, x_n)$ and ${\bf{y}}=(y_1, \dots, y_n)$ in $GF(2)^n$ is
${\bf{x}}\cdot{\bf{y}}=\sum_{i=1}^n x_i y_i$. The {\em dual} of $C$,
denoted by $C^{\perp}$ is the set of vectors orthogonal to every
codeword of $C$ under the Euclidean inner product. If $C=C^{\perp}$,
$C$ is called {\em self-dual}. A self-dual code is called {\em Type
II ({\mbox{or}} doubly-even)} if every codeword has weight divisible
by $4$, and {\em Type I ({\mbox{or}} singly-even)} if there exists a
codeword whose weight is congruent to $2 \pmod{4}$.

Two codes over $GF(2)$ are said to be {\em equivalent}
 if they differ only by a permutation
of the coordinates. Let $\C$ be a binary self-dual code of length
$n$ and minimum distance $d(C)$. Then $d(C)$ satisfies the
following (see~\cite{RS}).

$$d(C)  \le \left\{
\begin{array}{ll}
4  \left[\frac{n}{24}\right] +4 , & \mbox{ if } n \ne 22 \pmod{24}, \\
4  \left[\frac{n}{24}\right] +6 ,& \mbox{ if } n  = 22 \pmod{24}.
\end{array}\right.$$
A self-dual code meeting one of the above bounds is called {\em
extremal}. A code is called {\it optimal} if it has the highest
possible minimum distance for its length and dimension.

By the well known Gleason's theorem, the weight enumerator
$W_C(x,y)$ of a Type~I code can be written as follows (for rational coefficients $c_i$):
$$W_C(x,y)=\sum_{i=0}^{[n/8]} c_i(x^2+y^2)^{\frac{n}{2}-4i}\{x^2y^2(x^2-y^2)^2\}^i.$$

An important notion associated to a Type~I code is the {\em shadow}
$S$ of a code $C$, defined by $S=C_0^{\perp} \backslash C$, where $C_0$ is
the doubly-even subcode of $C$. In \cite{CS}, Conway and Sloane show
that for a weight enumerator $W_C(x,y)$ given above, the weight
enumerator $W_S$ of $S$ satisfies

$$W_S(x,y)=\sum_{i=0}^{[n/8]} c_i(-1)^i2^{\frac{n}{2}-6i}(xy)^{\frac{n}{2}-4i}(x^4-y^4)^{
2i}.$$

This notion of shadow permits to give more information on potential
weight enumerators of self-dual codes, and is also used to define
$s$-extremal codes (see \cite{BG} or Sec.~\ref{sec:s-ext}).

\medskip

The main tool to classify self-dual codes is based on the so-called
mass formula. It is known from \cite{Ple-F2} that self-dual binary
codes (Type~I or Type~II) of length $n$ satisfy a formula (a mass
formula):

$$N(n)=\sum_{\substack{j}} \frac{n!}{|Aut(C_j)|},$$
where the sum is made over all inequivalent self-dual codes (Type~I
or Type~II) of length $n$, $|Aut(C)|$ denotes the order of the
automorphism group of a code $C$, and $N(n)$ is the number of
Type~I or Type~II codes. In particular, for Type~I codes, $N(n)
=\prod_{i=1}^{\frac{n}{2}-1}(2^i+1)$ and for Type~II codes
$N(n)=\prod_{i=0}^{\frac{n}{2}-2}(2^i+1)$.

Therefore, for $n=38$,

\[
N(38)=\prod_{i=1}^{18}(2^i+1)= \sum_{\substack{j}}
 \frac{38!}{|Aut(C_j)|}.
 \]

 Hence,
\[
\begin{array}{lll}
13644432.20346 &< & \frac{\prod_{i=1}^{18}(2^i+1)}{38!} \\
 & = &
\sum_{\substack{j}}
 \frac{1}{|Aut(C_j)|} \\
  & \le & \#({\mbox{all inequivalent self-dual codes)}} \\
  \end{array}
  \]

Moreover, as there is no mass formula for extremal
self-dual codes, it might be also difficult to classify all extremal
binary $[38, 19, 8]$ codes. However, using the recursive
construction~\cite{AG} which was used in classifying all
extremal binary $[36, 18, 8]$ codes, we are successful in
classifying all extremal binary $[38, 19, 8]$ codes.

\medskip

A very interesting tool for self-dual codes is the subtraction procedure
of (11) on two coordinates of a code.
This procedure permits to construct a $[2n,n,d' \ge d]$ self-dual
code from a $[2n+2,n+1,d+2]$ self-dual code. It works as follows:
suppose one starts from a $[2n+2,n+1,d+2]$ self-dual code $C$ for $d \ge 2$.
Let $i$ and $j$ be two different coordinates of the columns of $C$. Since $d+2 \ge 3$
and $C$ is self-dual, any two columns of $C$ are independent (if not, there should be a codeword of weight $2$ in $C$, a contradiction).
This implies that the coordinates of the two columns of the codewords of $C$ contain $(00),(10),(01)$ and $(00)$.
For the subtraction procedure of $(11)$ on columns $i$ and $j$, one first keeps
all codewords which are either $(00)$ or $(11)$ on columns $i$ and $j$, and then
 deletes columns $i$ and $j$ for these codewords. Let $C'$ be the obtained code.
Since $d+2 > 2$ and by an argument similar to the shortening of a code, the dimension of $C'$ is $n$. Moreover since the scalar product
of any two codewords of $C$ is $0$, the scalar product of any two codewords of
$C'$ is also $0$. Now as the minimum distance of $C$ is $d+2$, the minimum distance
$d'$ of $C'$ is either $d$ either $d+2$ (depending on the fact that columns $i$ and $j$
intersect or not with codewords of $C$ of weight $d+2$). Overall
$C'$ is a $[2n,n,d'\ge d]$ self-dual code.

\medskip

\section{Construction methods}

There exist several methods to construct self-dual codes of length
$n+2$ from self-dual codes of length $n$.
In this section we recall these methods;
the recursive construction, the building-up construction and the
Harada-Munemasa construction. We eventually compare them.

\subsection{ The recursive construction}
\label{subsec:A}

 In~\cite{AG}, Aguilar and Gaborit give a recursive
construction of binary self-dual codes.
This algorithm can be seen as the reverse operation of the subtraction procedure
of (11) given above.  We recall that a subtraction procedure produces a self-dual $[2n,n,d'\ge d]$ code $C'$ from  a self-dual $[2n+2,n+1,d+2]$ code
$C$.
The recursive algorithm starts from a self-dual $[2n,n,d]$ code $C'$ and constructs
(up to permutation) all self-dual $[2n+2,n+1,d+2]$ codes which by subtraction
of (11) on certain two columns give the code $C'$. The idea of the recursive algorithm is very simple and consists of extending the code $C'$ with $11$ for all codewords of weight $d$, then constructing
all possibilities with $(00)$ or $(11)$ for a basis of remaining codewords, and eventually
checking for addition of a vector strictly contained in the shadow of the extended code.
This approach is very useful in classifying extremal self-dual $[2n+2,n+1,d+2]$ codes
because it is sufficient to know (up to permutation) a classification of $[2n,n,\ge d]$
self-dual codes. Indeed, any $[2n+2,n+1,d+2]$ code gives a $[2n,n,d]$ code
by subtraction of $(11)$ on adequate columns,
conversely applying the `reverse subtraction' procedure to the set of all
$[2n,n,d]$ codes (up to permutation) permits to construct a set of codes
which contains (up to permutation) all $[2n+2,n+1,d+2]$ codes.

We now recall the recursive algorithm (with a correction of $n-k$ in Step 2) from \cite{AG} into $\frac{n}{2}-k$:

\noindent \rule[-.1cm]{\linewidth}{0.2mm}
{\bf Recursive algorithm}\\
\noindent \rule[.3cm]{\linewidth}{0.2mm}

\vspace{-.3cm}
\noindent\textbf{Input:} $S_n$, the set of  $[n,\frac{n}{2},d]$ self-dual codes up to permutation\\
\textbf{Output:} The set of $[n+2,\frac{n}{2}+1,d+2]$ self-dual codes

\vspace{.2cm}
\noindent For each code $C_n$ of $S_n$ do:\\
\begin{small}
\begin{enumerate}
\item List all the words of weight $d$ and
construct the subcode $C_d$ of dimension $k$ generated by these
words. Construct a generator matrix $G_d$ of $C_d$ composed only
with words of weight $d$.
\item Let $E$ be a code of dimension $\frac{n}{2}-k$
with generator matrix $G_E$ such that
$C_n=C_d + E$, constructs the extended codes $C$ with generator
matrices:
\begin{equation} \label{eq:recursive}
\left[\begin{array}{ccccc}
1&1& & &\\
\vdots&\vdots& & G_d &\\
1&1& & &\\
a_1&a_1& & &\\
\vdots&\vdots& &G_E&\\
a_{\frac{n}{2}-k}&a_{\frac{n}{2}-k}&  & &\\
\end{array}\right]
\end{equation}

such that $a_i \in \{0,1\}, (1 \le i \le \frac{n}{2}-k).$
\item Complete all the previous codes ${C}$ by nonzero
elements of ${C}^{\perp}/C$ in order to obtain a self-dual code $D$
and check for codes with minimum distance $d+2$. For codes with
weight $d+2$ check for the equivalence with already obtained
self-dual $[n+2,\frac{n}{2}+1,d+2]$ codes.
\end{enumerate}
\end{small}
\noindent \rule[.3cm]{\linewidth}{0.2mm}

The main result of \cite{AG} is the following:

\medskip

\begin{theorem} Applying the previous recursive algorithm
to the set of all inequivalent (up to permutation) binary self-dual
$[n,n/2,d]$ codes permits to find all inequivalent self-dual binary
$[n+2,n/2+1,d+2]$ codes.
\end{theorem}

\medskip

\subsection{The building-up construction}

 There are other constructions generating self-dual codes of
length $n+2$ from self-dual codes of length $n$. In particular, we
compare the above construction with two constructions; the
building-up construction~\cite{Kim} by Kim, and Harada-Munemasa's
construction~\cite{HarMun_10} since both constructions generate all
self-dual codes of length $n+2$ from the set of all self-dual codes
of length $n$.

\medskip

\begin{theorem}~(\cite[building-up]{Kim})
\label{kimthm} Let $G_0=({\bf{r_i}})$ be a generator matrix (may not
be in standard form) of a self-dual code $\C_0$ over $GF(2)$ of
length $n$, where
 ${\bf{r}}_i$ is a row of $G_0$ for $1 \le i \le n/2$.
Let $\mbox{\bf x}$ be a vector in $GF(2)^{n}$ with an odd weight.
Define $y_i:={\bf{x}} \cdot {\bf{r}}_i$ for $ 1 \le i \le n/2$,
where $\cdot$ denotes the usual inner product. Then the following
matrix

\begin{equation} \label{eq:build-up}
G = \left[ \begin{array}{cc|cccccc}
  1   & 0   & &  &  & {\bf{x}}   &  &  \\ \hline
  y_1 & y_1 &    &        &     &         &        &        \\
  \vdots & \vdots &  &  & & G_0  & & \\
 y_{n/2} & y_{n/2} &    &        &     & \\
\end{array}
\right]
\end{equation}

 generates a self-dual code $C$ over $GF(2)$ of length $n+2$.
\end{theorem}

The converse of the building-up construction holds as follows.
\medskip

\begin{theorem}(\cite{Kim})
\label{one_to_one} Any self-dual code $C$ over $GF(2)$ of length $n$
with minimum weight $d > 2$ is obtained from some self-dual code
$C_0$ of length $n-2$ (up to equivalence) by the construction in
Theorem~\ref{kimthm}.
\end{theorem}

\medskip

The recursive construction is a special case of the building-up
construction. The reason is as follows.

We show that the matrix in the form~(\ref{eq:recursive}) together
with a representative in $C^{\perp}\slash C$ whose weight is $> 2$
can be written in the form~(\ref{eq:build-up}) up to permutation
equivalence. Suppose we are given the matrix in the
form~(\ref{eq:recursive}) above and let $C$ be the code generated by
this matrix. Then there are four cosets of $C$ in $C^{\perp}$; that
is, $C$, ${\bf{z}}_1+C$, ${\bf{z}}_2+C$, and ${\bf{z}}_1+
{\bf{z}}_2+C$ for some nonzeroes ${\bf{z}}_1, {\bf{z}}_2 \in
GF(2)^{n+2}$. We may assume that ${\bf{z}}_1= (1, 1, 0, 0, \cdots,
0)$ since ${\bf{z}}_1$ is nonzero and orthogonal to $C$. Then the
minimum weight of $C \cup ({\bf{z}}_1+C)$ is $2$, which is excluded.
Hence by permuting the first two columns of ${\bf{z}}_2$ if needed,
we may put ${\bf{z}}_2= (1, 0 ~|~ {\bf{x}})$ where ${\bf{x}} \in
GF(2)^n$. As $C \cup ({\bf{z}}_2+C)$ is designed to be self-dual,
${\bf{z}}_2$ is orthogonal to itself; hence ${\bf{x}}$ is odd. Then
as ${\bf{z}}_2 \cdot (1, 1~|~{\bf{r}}_i) =0$, where ${\bf{r}}_i$ is
a row of $G_d$ in the form~(\ref{eq:recursive}) for $1 \le i \le k$,
we have ${\bf{x}} \cdot {\bf{r}}_i =1$. Thus by letting
$y_i:={\bf{x}} \cdot {\bf{r}}_i =1$ for $1 \le i \le k$, we obtain
the matrix of the form~(\ref{eq:build-up}). This implies that the
recursive construction is a special case of the building-up
construction.

\medskip

\subsection{The Harada-Munemasa construction}
In what follows, we recall Harada-Munemasa's
construction~\cite{HarMun_10}. We note that this is a binary version
of Huffman's construction~\cite{Huf_97} for Hermitian self-dual
codes over $GF(4)$.

Let $G_1$ be a generator matrix of a self-dual $[n, n/2, d]$ code
$C_1$. Then the matrix
\begin{equation} \label{eq:HarMun}
G_2:=\left[\begin{array}{ccccc}
a_1& a_1& & &\\
\vdots&\vdots& & G_1 &\\
a_{n/2 -1}& a_{n/2 -1}& & &\\
\end{array}\right],
\end{equation}
where $a_i \in GF(2)$ for $(1 \le i \le n/2-1)$, generates a
self-orthogonal $[n+2, n/2]$ code $C_2$. The matrix of the
form~(\ref{eq:HarMun}) is a general form of~(\ref{eq:recursive}) in the
recursive construction. In order to reduce the possibilities of
$a_i$'s, they~\cite{HarMun_10} consider the orbits of the vector
$a^T:=(a_1, \cdots, a_{n/2 -1})^T$ under a certain subgroup of
$GL(n/2 -1, 2)$ to get equivalent self-dual codes of length $n+2$.
After reducing the possibilities, as in the recursive construction,
add to $C_2$ a coset ${\bf{z}}_2 +C_2$ from $C_2^{\perp}\slash C_2$
whose weight is $> 2$ to get a self-dual $[n+2, n/2+1, d'>2]$ code.
Unlike the recursive construction, Harada-Munemasa's construction
does not necessarily give self-dual $[n+2, n/2+1]$ codes with
minimum weight $d'=d+2$.

%This recursive construction is
%similar to the building-up construction~\cite{Kim} an
%Harada-Munemasa's construction~\cite{HarMun_10} (see also Huffman's
%construction~\cite{Huf_97}). However, the recursive construction is
%more efficient in constructing many self-dual codes with higher
%minimum distance because of a small number of possibilities of
%$a_i's$ in Step $2)$.

\subsection{Comparison of the different methods}

The recursive construction is specially interesting when one wants
to classify extremal codes since it permits to obtain a partial
classification for a given minimum distance while other
constructions do need to start from a whole classification.

More precisely, the recursive construction is more efficient than
the building-up construction in generating many self-dual codes with
higher minimum weight. This is because the recursive construction
checks a relatively small number of possibilities of $a_i's$ in Step
$2)$, whose complexity is $2^{n/2 -k}$, where $k \ge 1$ depends on
the given code. From our experimental results, the dimensions $k$ of
subcodes of the $58671~[36,18,6]$ codes  generated by linearly
independent vectors of weight $6$ lie between $2$ and $18$. We give
the possible values of $k$ and the number ${\mbox{num}}$
 of their subcodes in Table~\ref{tab:sub_36}.

\begin{table} [thb]
\centering
\caption{Number of self-dual $[36,18,6]$ codes whose subcode generated by codewords of weight $6$ has dimension $k$}
\label{tab:sub_36}
\begin{tabular}{c|c||c|c||c|c|}
${\mbox{dim}}~k$& {\mbox{num}}& ${\mbox{dim}}~k$& {\mbox{num}}&
 ${\mbox{dim}}~k$& {\mbox{num}} \\
\hline
2  & 148  & 8& 4615 & 14 & 8170   \\
3  & 5    & 9 & 911  & 15 & 5311  \\
4  & 666  & 10 & 7165 &16 & 6290  \\
5  & 45   & 11 & 2299  & 17 & 4492 \\
6  & 2165 & 12 & 8411 & 18 & 3615 \\
7  & 263  & 13 & 4100 &  &  \\
\end{tabular}
\end{table}

We see from our table that there are much more subcodes of large
dimension than those of small dimension and this clearly shows the
efficiency of our recursive algorithm.

On the other hand, the
building-up construction~\cite{Kim} needs $2^{n-1}$ possibilities
for the choice of odd vectors ${\bf{x}}$, generating all self-dual
codes with various minimum distances. This complexity can be reduced
to $2^{n/2}$ as remarked in~\cite{GulHarKim}, which is still higher
than that of the recursive construction.

As described above, Harada-Munemasa's construction is effective if
the given code has a large automorphism group in order to reduce the
complexity of checking the equivalence. For example, if $n=36$, then
$41019$ (respectively $11242$) out of the $58671$ self-dual
$[36,18,6]$ codes~\cite{HarMun_10} have the automorphism group order
$1$ (respectively $2$). Thus Harada-Munemasa's construction usually
requires $2^{19}$ or $2^{18}$ possibilities to generate self-dual
codes of length $38$ with various minimum distances, given a
$[36,18,6]$ self-dual code.

Overall, we conclude that when we classify binary self-dual $[38,
19,8]$ codes, the recursive algorithm is much faster than the other
two constructions.

\section{Classification of the $[38,19,8]$ self-dual codes }

\subsection{Construction of all $[38,19,8]$ self-dual codes}

There are two possible weight enumerators $W_1, W_2$ and shadow
weight enumerators $S_1, S_2$ for an extremal self-dual $[38, 19,
8]$ code~\cite{CS}.

\begin{align}
W_1 &= 1+ 171 y^8 + 1862 y^{10} + \cdots  \label{W1} \\
S_1 &=  114 y^7 + 9044 y^{11} + 118446 y^{15} + \cdots; \label{S1} \\
W_2 &=  1 + 203 y^8 + 1702 y^{10} + \cdots \label{W2}\\
S_2 &= y^3 + 106 y^7 + 9072 y^{11} + 118390 y^{15}
\end{align}

In~\cite{CS} two self-dual $[38,19,8]$ codes with $W_1$, denoted by $R_3$ and $D_4$, were given, where $|{\mbox{Aut}}(R_3)|=1$ and $|{\mbox{Aut}}(D_4)|=342$. In~\cite{HarKim} one self-dual $[38,19,8]$ code $C_{38}$ with $W_2$ was given with $|{\mbox{Aut}}(C_{38})|=1$.
Then Harada~\cite{Har} gave $40$ self-dual $[38,19,8]$ codes with $W_1$ and $W_2$ and automorphism group orders $1, 2, 4, 8$. Later, Kim~\cite{Kim} constructed $325$ self-dual $[38,19,8]$ codes with $W_1$ and $W_2$ and automorphism group orders $1, 2, 3$. Hence there are at least $368$ inequivalent self-dual $[38,19,8]$ codes.
We show that there are exactly $2744$ inequivalent self-dual $[38,19,8]$ codes.

Starting from the 58671 $[36,18,6]$ codes of \cite{HarMun_10}, we
apply the recursive algorithm of Section~\ref{subsec:A}.
The more expensive part of the algorithm is the inequivalence
testing of the differently constructed codes. In order to optimize the
computation we separated the 58671 $[36,18,6]$ codes into sets
$S_{36,i}$ of $1000$ codes. To each set, we apply the recursive
algorithm to obtain a list $S_{38,i}$ of inequivalent $[38,19,8]$
codes derived from the set $S_{36,i}$. Each set $S_{38,i}$ contains
a number of inequivalent codes. Then we compared all
the $S_{38,i}$ sets to eventually obtained a list of all
inequivalent $[38,19,8]$ self-dual codes. This method permits to avoid many
costly inequivalence comparisons between codes, since separating the
whole list of $[36,18,6]$ codes permits to avoid inequivalence
testing as the $S_{38,i}$ list starts from an empty list.

The whole process took about three weeks on a CPU
2.53GHz computer.

Now we obtain our main theorem below.

\medskip

\begin{theorem}
There are exactly $2744$ inequivalent extremal self-dual $[38,19,8]$ codes.
\end{theorem}

\medskip

In Table~\ref{tab:ext_38}, we describe all extremal self-dual $[38,19,8]$ codes with respect to their orders, where $|{\mbox{Aut}}(C)|$ and {\mbox{num}} stand for the order of automorphism group
and the number of codes respectively.

\begin{table} [thb]
\centering \caption{Number of extremal self-dual $[38, 19, 8]$ codes with respect to their orders}
\label{tab:ext_38}
\begin{tabular}{c|c||c|c||c|c|}
$|{\mbox{Aut}}(C)|$& {\mbox{num}}& $|{\mbox{Aut}}(C)|$& {\mbox{num}}&
 $|{\mbox{Aut}}(C)|$& {\mbox{num}} \\
\hline
1  & 2253  & 9 & 1 & 36 & 1   \\
2  & 322 & 12 & 8  & 144 & 1  \\
3  & 36  & 14 &  1 &168 & 2  \\
4  & 68  & 18 & 1  & 216 & 1 \\
6  &17   &  21 & 1 & 342 & 1 \\
8  & 15  & 24 &14 & 504 & 1 \\
\end{tabular}
\end{table}

As mentioned above, the previously known self-dual $[38, 19, 8]$
codes have automorphism group orders $1,2,3,4,8,$ and $342$. Hence
we list several new self-dual $[38, 19, 8]$ codes $C_{38}^i$ with
different automorphism group orders $|{\mbox{Aut}}(C_{38}^i)|=i=6,
9, 12, 14, 18, 21, 24, 36, 144, 168, 216, 504$ in Appendix. To
save space, we only give one code for each order. We also list
$C_{38}^{342}$ which is equivalent to the double-circulant code
$D_4$ in~\cite{CS}.  The list of all extremal self-dual
$[38,19,8]$ codes can be obtained at
http://www.unilim.fr/pages$\_$perso/philippe.gaborit/SD/GF2/GF2I.htm
.

%As some new examples, we give the generator matrices $G(C_{38}^i)$ of the codes $C_{38}^i$ with the automorphism group order $|{\mbox{Aut}}(C_{38}^i)|=i$. To save space, we only give one code for each order. We note that the double-circulant code $D_4$ in~\cite{CS} is equivalent to $C_{38}^{342}$ since both have the same automorphism group order $342$.

\subsection{ An up-to-date table of the number of classified optimal self-dual codes}

In the following we give an up-to-date table of the classification of optimal Type I self-dual codes, where
being optimal means that this is the best possible minimum distance among self-dual codes of a given length. These
codes may not be extremal in the classical sense. For instance, an extremal
self-dual code of length $34$ will have minimum distance $8$ if exists, but it is known that such a code cannot exist and the optimal minimum distance is $6$.
The highest length (up to now) for which Type I optimal codes are classified is length $38$, which is done in this paper for the first time.
Notice that it is length $48$ for Type II codes. Complete references for the self-dual codes
can be found for instance in \cite{Huffman} and \cite{NRS}, except for length $38$.

\begin{table}
\centering \caption{Number of optimal Type I and Type II codes}
\begin{tabular}{c|c|c||c|c|c}
$n$&$d$&$\mbox{num}$& $n$&$d$&$\mbox{num}$\\
\hline
2  & 2 & 1 & 22 & 6  &1 \\
4  & 2 & 1 & 24 & 8 &1  \\
6  & 2 & 1 & 26 & 6 & 1 \\
8  & 4 & 1 & 28 & 6 & 3 \\
10  &2  &1  & 30 & 6 &13  \\
12  & 4 &1  & 32 & 8  &8 \\
14  & 4 &1  & 34 & 6 &938  \\
16  & 4 &3  & 36 & 8 &  41\\
18  & 4 &2  & 38 & 8 & 2744\\
20  &4  &7 &   &  &  \\
\end{tabular}
\end{table}

\section{Classification of $s$-extremal codes }
\label{sec:s-ext}

In this section, we classify $s$-extremal codes of length $38$ and $d=8$ together with
$s$-extremal codes of length $38$ and $d=6$ .

\subsection{$s$-extremal codes}

The notion of $s$-extremal codes was introduced by Bachoc and
Gaborit in \cite{BG}. This
type of codes is related to the notion of self-dual codes with long shadows introduced by Elkies in \cite{E95}.
We recall the definition of $s$-extremal codes from~\cite{BG}.

Let $C$ be a Type I self-dual binary code of length $n$. We denote by $C_0$ the doubly-even
subcode of $C$. We denote by $\x$ an element of $C \backslash C_0$. The shadow $S$ is defined by
$S=C_0^{\perp} \backslash C$, we denote by $\y$ an element of $S \backslash C$.
We have $C_0^{\perp}=C_0 \cup C_1 \cup C_2 \cup C_3$
for $C_1=\y+ C_0, C_2=\x+C_0$ and $C_3=\x+\y+C_0$.
Then it is well known that $C=C_0 \cup C_2$ and $S=C_1 \cup C_3$.
We have moreover the following three facts \cite{CS}:
\begin{enumerate}
\item for any $\y \in S$, ${\mbox{weight}}(\y) \equiv \frac{n}{2} \pmod 4$
\item for any $\y \in S$ and $\x \in C_2 : \x \cdot \y=1$,
\item for any $\y \in S$ and $\z \in C_0 : \x \cdot \z=0$.
\end{enumerate}

We denote the weight
enumerators of $C$ and $S$ by $W_C$ and $W_S$, respectively. From \cite{CS}, there exist
$c_0, \dots,c_{[n/8]}\in \R$ such that:

\begin{equation}\label{E1}
\bf \begin{cases}
W_C(x,y)&=\sum_{i=0}^{[n/8]} c_i(x^2+y^2)^{\frac{n}{2}-4i}\{x^2y^2(x^2-y^2)^2\}^i\\
W_S(x,y)&=\sum_{i=0}^{[n/8]} c_i(-1)^i2^{\frac{n}{2}-6i}(xy)^{\frac{n}{2}-4i}(x^4-y^4)^{
2i}
\end{cases}
\end{equation}

 Let $d$ be the minimum weight of $C$ and $s$ the minimum weight of its shadow.

\begin{theorem}\cite{BG} Let $C$ be a Type I self-dual binary code of length $n$ with minimum weight $d$, and let $S$ be its shadow with minimum weight $s$. Then,
$2d+s \le 4+\frac{n}{2}$, unless $n \equiv 22 \mod 24$ and $d=4[n/24]+6$,
in which case $2d+s = 8+\frac{n}{2}$.
\end{theorem}

\medskip

 A Type I code whose parameters $(d,s)$ satisfy the equality in the previous bounds is called {\em $s$-extremal}. In that case, the
  polynomials $W_C$ and $W_S$ are uniquely
determined.

A bound for $n$ when the minimum weight $d$ of an $s$-extremal code is
divisible by $4$ has been given in \cite{G} and in \cite{HanKim},
and a bound has also been given for $d=6$~\cite[Theorem 4.1]{BG} and
$d \equiv 2 \pmod{4}$ with $d >6$ \cite{HanKim}.

\medskip

\begin{theorem}{\rm(\cite{G}, \cite{HanKim})}

Let $C$ be an $s$-extremal code with parameters $(s,d)$ of length
$n$. If $d \equiv 0 \pmod 4 $, then $n < 6d-2$.

\end{theorem}

\medskip

\begin{theorem}{\rm(\cite{HanKim})}

Let $C$ be an $s$-extremal code with parameters $(s,d)$ of length
$n$. If $d > 6$ and $d \equiv 2 \pmod 4 $, then $n < 21d-82$.

\end{theorem}

\medskip

Before proving our classification of $s$-extremal codes of length $38$, we prove a result which permits in certain
cases to relate the weight of the shadow of a code $C$ with the weight of the shadow
of a subtracted code by (11):

\medskip

\begin{theorem} \label{thm:s-ext_short}
If $C$ is a $[4n+2,2n+1,d+2]$ self-dual code with $d \equiv 0 \pmod 4$, $d \ne 0$ and shadow weight $s \ge 3$,
then there exist two coordinates of $C$ on which the subtraction of (11)
gives a self-dual $[4n, 2n, d]$ code $C'$ with shadow weight $s-1$.
\end{theorem}

\medskip

\pf Our proof is based on the existence of the following four vectors $\x,\y,\z$, and ${\bf{t}}$ such that:
\begin{enumerate}
\item $\y=(\y'10)$, $\y \in S$ of weight $s$
\item $\x=(\x'11)$, $\x \in C_2$ of weight $d+2$
\item $\z=(\z'11)$, $\z \in C_0$
\item ${\bf{t}}=({\bf{t}}'10)$, ${\bf{t}} \in C_0$
\end{enumerate}

Let $\y \in S$ of weight $s$ and $\x  \in C_2$ of weight $d+2$. We have $\x \cdot \y=1$, that is, $\x$ and $\y$ meet in an odd number of positions. Then $x_i=1=y_i$ for some $i$. As the weight of $\x$ is even, there is a $j$ such that $x_j=1$ and $y_j=0$. Up to permutation, we may assume that $\x=(\x'11)$ and $\y=(\y'10)$. Now it remains to show that there exist $\z$ and ${\bf{t}}$ given above. To do this, note that $C_0^{\perp}=C \cup S$. Hence the minimum distance of $C_0^{\perp} =\min\{d+2, s\} \ge 3$.
Hence every two columns of a generator matrix of $C_0$ are linearly independent. (This means that $C_0$ has strength $2$. See \cite[p. 435]{HufPle} for the term.)
Thus in each set of two columns of $C_0$ each binary $2$-tuple occurs the same number $|C_0|/4$ of times. Therefore there exist $\z=(\z'11) \in C_0$ and ${\bf{t}}=({\bf{t}}'10) \in C_0$.

Since the coordinates of $\z$ and ${\bf{t}}$ are respectively
(11) and (10) on the last two positions, there exists a doubly-even code $C_0''$
of dimension $2n-2$ such that the doubly-even subcode
$C_0$ of $C$ can be written:
$$
\left(
\begin{array}{lll}
\z'  & 1  & 1    \\
{\bf{t}}'  & 1 & 0   \\
\hline
  & 0  & 0   \\
C_0'' & \vdots & \vdots   \\
  &0 &  0
\end{array}
\right)
$$
 Now if one subtracts (11) on the two last columns of $C$
one obtains a code $C'$, such that its doubly-even subcode $C_0'$
has dimension $2n-1$, ($2n-2$ vectors of $C_0''$ and the vector $\x'$ - which cannot
be null since $d+2 \ne 2$), the subcode $C_0'$ can be written as:

$$
\left(
\begin{array}{l}
\x'     \\
C_0''
\end{array}
\right)
$$
Overall a generator matrix of $C'$ can be written as:

$$
\left(
\begin{array}{l}
\z'   \\
\x'     \\
C_0''
\end{array}
\right)
$$
with $\x'$ of weight $d$. And $C_2'=C_0'+\z'$.
Let $\cc'$ be in $C_0''$ and denote by $\cc$ the extension of $\cc'$ with (00), then
$\cc \in C_0$. Now $\y' \cdot \cc'=0$ since $\y \cdot \cc=0$ and $\y' \cdot \x'=0$ since $\y \cdot \x=1$,
which proves that for $\cc \in C_0'$, $\cc \cdot \y'=0$. Moreover since
$\y \cdot \z=0$, we deduce that $\y' \cdot \z'=1$.
The latter results show that $C'$ is a $[4n,2n]$ self-dual code
with minimum distance $d$ (since $\x'$ has weight $d$), such that $C'=C_0' \cup (\z'+C_0')$
and with shadow $S'=(\y'+C_0') \cup  (\y'+\z'+C_0')$.
Finally we remark that by construction, for any vector of $S'$
it is possible to add either $(11),(01),(10),$ or $(00)$
such that the extended vector is in $S$. Since all the weights
of $S'$ are congruent to $s-1 \pmod 4$ and since $\y'$ of weight
$s-1$ is in $S'$ we deduce that the minimum weight of $S'$ is $s-1$
which proves the theorem.
\qed

\subsection{Classification of $s$-extremal $[38,19,8]$ codes}
\medskip

Let $C$ be an extremal self-dual $[38, 19, 8]$ code. If $C$
satisfies $W_1$ in equation~(\ref{W1}), then $S_1$ in~(\ref{S1}) is
also satisfied. So we have $d=8$ and $s=7$; hence $2d+s=23=n/2+4$.
This implies that $C$ is an $s$-extremal code with parameters
$(7,8)$. Clearly if $C$ satisfies $W_2$, then $C$ cannot be an
$s$-extremal code since $2d+s=19 < 23$. The $s$-extremal code can be obtained
directly from the classification of all $[38,19,8]$ by a simple
computation on the weight enumerator. We obtain:

\medskip

\begin{theorem} \label{thm_s-ext-38-8}
There are exactly $1730$ $s$-extremal $[38,19,8]$ codes.
\end{theorem}

\subsection{Classification of $s$-extremal $[38,19,6]$ codes}\label{sec:long}

The case of $d=6$ was mainly considered in \cite{BG}, where $s$-extremal codes are known to exist for the lengths $22 \le n \le 44$. Two lengths $36$ and $38$
remained open in~\cite{BG}.  Later, $s$-extremal codes of length $36$ and $d=6$ were classified in \cite{AG}.
The only open case is the classification of $s$-extremal codes of length $38$ and $d=6$. There are at least two such codes as shown in~\cite{BG}. We show that there are exactly two $s$-extremal codes of length $38$ and $d=6$.

For a self-dual $[38, 19, 6]$ code to be $s$-extremal, the minimum weight of its shadow must be $s=11$.
A simple approach to find all $s$-extremal $[38,19,6]$ codes
is to apply the recursive construction, starting from the set of
all inequivalent $[36,18,4]$ self-dual codes. Unfortunately, since
there are $436,633$ $[36,18,4]$ self-dual codes, such a computation would require
more than $80$ days, and although it is doable theoretically, in practice it
remains largely too costly. Fortunately, by using the fact that
such an $s$-extremal code has a shadow with high minimum weight
it is possible to dramatically decrease this computation.

We have shown in Theorem~\ref{thm:s-ext_short} that it is possible to relate
the weight of a shadow
of a code to the that of the shadow of the subtracted code under certain
conditions. We use this result to prove the following classification theorem:

\begin{theorem} \label{thm_s-ext-38-6}
There are exactly two $s$-extremal $[38,19,6]$ codes.
\end{theorem}
\pf Let $C$ be an $s$-extremal  $[38,19,6]$ code, then $C$ has shadow weight $s=11$.
Applying Theorem~\ref{thm:s-ext_short} we deduce that there exist two coordinates on which the subtraction of $(11)$ of $C$ produces a $[36,18,4]$ self-dual code with shadow weight
$10$. Hence if one applies the recursive algorithm starting from the set $S_{36,10}$ of all
inequivalent $[36,18,4]$ self-dual codes with shadow weight 10,
we construct the set of all $[38,19,6]$ self-dual
codes (up to permutation) which by a subtraction of (11) on certain two columns give the set $S_{36,10}$.
Hence applying the recursive algorithm to $S_{36,10}$ gives a set of self-dual codes which contains
all $s$-extremal $[38,19,6]$ codes.
In practice, from the classification of \cite{HarMun_10}, there are exactly $24$
$[36,18,4]$ self-dual codes with shadow weight $10$. The application of the recursive
algorithm is then fast with these codes and we have that there are exactly
two $s$-extremal $[38,19,6]$ codes.
\qed

The two $s$-extremal codes $C_{38,1}, C_{38,2}$ have covering radius $11$ and their generator matrices $G(C_{38,1}), G(C_{38,2})$ are as follows:
{\tiny
\[
G(C_{38,1})=\left[\begin{array}{c}

       10000000000000000000000011111100001110 \\
       01000000000000000000000011111100000001 \\
       00100000000000000001010101010101111001 \\
       00010000000000000001010101010101000101 \\
       00001000000000000000000101010101110011 \\
       00000100000000000000000101010110111111 \\
       00000010000001000000000001111000000000 \\
       00000001000001000000000001110111000000 \\
       00000000100001010000000110100011000000 \\
       00000000010001010000000110010000000000 \\
       00000000001001010000010000100101010110 \\
       00000000000101010000010011011010010110 \\
       00000000000011000000000011000011000000 \\
       00000000000000110000000011001100000000 \\
       00000000000000001001000101010110011001 \\
       00000000000000000101000110101001101001 \\
       00000000000000000011000011111111001100 \\
       00000000000000000000110011111111111100 \\
       00000000000000000000001111001111000000 \\
\end{array}\right]
\]

\[
G(C_{38,2})=\left[\begin{array}{c}
       10000000000000000100010101010110010010 \\
       01000000000000000100010101010110011101 \\
       00100000000000000101000000000000100110 \\
       00010000000000000101000000000000011001 \\
       00001000000000000000000101010101000000 \\
       00000100000000000000000101010110111100 \\
       00000010000000010101010001011101110101 \\
       00000001000000010101010001010010110101 \\
       00000000100000000101010110000110110101 \\
       00000000010000000101010110110101110101 \\
       00000000001000000000000110101001110101 \\
       00000000000100000000000101010110000110 \\
       00000000000010010101010011100110110101 \\
       00000000000001010101010000100101110101 \\
       00000000000000110000000011001100000000 \\
       00000000000000001100000011111111110000 \\
       00000000000000000011000011111111111111 \\
       00000000000000000000110011111111001111 \\
       00000000000000000000001111001111000000 \\
\end{array}\right]
\]
}

Notice that these codes were already known from \cite{BG}, but it was not known whether there exist other codes.

\subsection{Up-to-date tables for $s$-extremal codes}

In the following we give up-to-date tables for $s$-extremal codes of minimum distance
$6$ and $8$:

$\bullet \: d=6$ \\
For this minimum distance, we know that there are exactly two $s$-extremal codes of length $38$ and $d=6$ from Theorem~\ref{thm_s-ext-38-6}. This was the only unknown case (see \cite{AG}, \cite{BG}). Now we complete the classification of
$s$-extremal codes of $d=6$ in Table~\ref{tab:s-ext-38-6}.

\begin{table}[thb]
\centering \caption{Number of $s$-extremal codes with $d=6$} \label{tab:s-ext-38-6}
\begin{tabular}{c|c||c|c}
$n$&$\mbox{num}$& $n$&$\mbox{num}$\\
\hline
22  & 1  & 34 & 17   \\
24  & 1 & 36 & 5   \\
26  & 1  & 38 &  2   \\
28  & 2  & 40 & 1   \\
30  &9 &  42 & 1  \\
32  & 19  & 44 &1
\end{tabular}
\end{table}

$\bullet \: d=8$  \\
In this case, $s$-extremal codes exist for $32 \le n \le 44$. More precisely, $s$-extremal codes of length $32$ were known from the classification of extremal self-dual codes
of length $32$, and $s$-extremal codes of length $36$ were done in \cite{AG}. We have completed the classification $s$-extremal codes of length $38$ and $d=8$ from Theorem~\ref{thm_s-ext-38-8}.
We list currently known codes for $d=8$ in Table~\ref{tab:s-ext-38-8}.

\begin{table}[thb]
\centering \caption{Number of $s$-extremal codes with $d=8$} \label{tab:s-ext-38-8}
\begin{tabular}[t]{c|c|c}
$n$&$\mbox{num}$&$\mbox{ref}$\\
\hline
32  & 3 & \cite{CS}  \\
36  & 25 & \cite{AG} \\
38  &  1730 & \text{this paper} \\
40  & $\ge 4$ & \cite{CS},\cite{BuYo}   \\
42  & $\ge 17$  & \cite{CS},\cite{Bu}  \\
44  & $\ge 1$ & \cite{CS}
\end{tabular}\hfil

\end{table}

\section{Covering radii of self-dual codes of length $38$}

The {\em covering radius $\rho(C)$} of a code $C$ is the smallest
integer $R$ such that spheres of radius $R$ around codewords cover
$\mathbb F_2^n$.

The following theorems give the lower and upper bound of $\rho(C)$
for a self-dual code over $GF(2)$.

\medskip

\begin{theorem}(\cite{CKMS}, Theorem 1) \label{thm:spherecovering}
Let $C$ be a self-dual code of length $n$ over $GF(2)$. Then $\sum
_{i=0}^{\rho(C)} {_n}C_i \ge 2^{n/2}$. More precisely, $\sum
_{2i \le \rho(C)} {_n}C_{2i} \ge 2^{(n/2) -1}$ and
$\sum
_{2i+1 \le \rho(C)} {_n}C_{2i+1} \ge 2^{(n/2) -1}$,
where ${_n}C_{i}$ means $n$ choose $i$.
\end{theorem}

\medskip

\begin{theorem}(\cite{CKMS},\cite{HufPle}, Delsarte's bound)\label{thm:Del}
Let $C$ be a self-dual code of length $n$ over $GF(2)$ and $s$ be
the number of distinct nonzero weights in $C$. Then $\rho(C) \le s$.
\end{theorem}

\medskip

By Theorem~\ref{thm:spherecovering}, any self-dual $[38,19]$ code has covering radius at least $6$. On the other hand,
the weight enumerators~(\ref{W1}) and~(\ref{W2}) of any self-dual $[38,19, 8]$ code has $13$ nonzero weights. Thus by Theorem~\ref{thm:Del}, the covering radius of any self-dual $[38,19,8]$ code is at most $13$. Combining both, we have $6 \le \rho(C) \le 13$ for any self-dual $[38,19,8]$ code $C$.

Using our classification of all self-dual $[38,19, 8]$ codes, we have the following.

%\begin{center}
%\begin{tabular}{c|c|c|c|c|c|c|c|c|c|}
%R(C) &5 &6 &7 &8 &9 &10 &11 &12 &13\\
%\hline
%num &0 &0 &2744 &0 &0 &0 &0 &0 &0\\
%\end{tabular}
%\end{center}

\medskip

\begin{theorem}
All $2744$ self-dual $[38, 19, 8]$ codes have covering radius $7$.
\end{theorem}

\medskip

\begin{remark}
If we choose a coset representative of weight $7$ and using it as a
vector ${\bf{x}}$ in Theorem~\ref{kimthm}, then the built code will
be an extremal self-dual $[40,20, 8]$ code. Hence for $n=38$, any extremal self-dual $[38,19,8]$ code can produce an extremal self-dual $[40,20,8]$ code
using the building-up construction. This is not always true for some
lengths (e.g. $n=24, n=32$).
\end{remark}

\medskip

\begin{proposition}\label{prop:cov}
Let $C_1$ be a self-dual code of length $n$ and covering radius
$\rho(C_1)$. Then any self-dual code $C_2$ of length $n+2$ obtained by the building-up construction (in particular, by the recursive algorithm) has covering
radius $\rho(C_2) \le \rho(C_1) + 2$.
\end{proposition}

\pf
Let $\rho(C_1) =r$.
We recall~\cite[Theorem 1.25.5]{HufPle} that the covering radius
$\rho(C_1)$ of a linear code $C_1$ with parity check matrix $H_1$ is
the smallest number $s$ such that every nonzero syndrome is a
combination of $s$ or fewer columns of $H_1$, and some syndrome
requires $s$ columns.  The generator matrix
$G_2$ of $C_2$ by the building-up construction is of the
form~(\ref{kimthm}). This $G_2$ is also a parity check matrix of
$C_2$ as $C_2$ is self-dual. Any syndrome ${\bf{u}} = [u_1, u_2, \cdots, u_{(n/2) +1}]^T$ with
respect to $G_2$ can be written as $u_1 [1, 0, \cdots, 0]^T + [0, u_2,
\cdots, u_{(n/2) +1}]^T$. Now $[0, u_2, \cdots, u_{(n/2) +1}]^T$ is a
linear combination of $r$ or fewer columns of $G_2$ as $G_1$ has covering radius $r$, and $[1, 0
,\cdots, 0]^T$ is the difference of the first columns of $G_2$
in the form~(\ref{kimthm}). Hence ${\bf{u}}$ is a linear combination
of at most $r+2$ columns of $G_2$. Thus $\rho(C_2) \le \rho(C_1) + 2$.
\qed

Using Proposition~\ref{prop:cov}, we have a better upper bound for the covering radius of a self-dual $[38, 19, 6]$ code than Delsarte's bound as follows.
\begin{cor}
The covering radius $\rho(C)$ of any self-dual $[38, 19, 6]$ code is $6 \le \rho(C) \le 12$.
\end{cor}

\pf
The lower bound is true for any even $[38, 19]$ code by Theorem~\ref{thm:spherecovering}. Delsarte's bound would imply $\rho(C) \le 15$. For a better upper bound,
we recall that any self-dual $[38, 19, 6]$ code can be constructed from a self-dual $[36, 18, 4]$ code by the recursive algorithm.
Since the covering radius of any self-dual $[36, 18, d=4,6]$
code is at most $10$~\cite{HarMun_10}, it follows that the
covering radius of any self-dual $[38, 19, 6]$ code is at most $12$ by Proposition \ref{prop:cov}.
\qed

\bigskip

\centerline{ ACKNOWLEDGEMENT }
\medskip
All the computations were done with the MAGMA system \cite{mag}.
While this work was under review, the
authors learned that independently, similar results have been obtained by others (see \cite{BetHarMun}, \cite{BouBou}).

\bigskip

\centerline{APPENDIX}

\medskip

Let $i=6, 9, 12, 14, 18, 21, 24, 36, 144, 168, 216, 342, 504$.
Then $G(C_{38}^i)$ represents a generator matrix of a new self-dual  $[38, 19, 8]$ code $C_{38}^i$ with the automorphism group order $|{\mbox{Aut}}(C_{38}^i)|=i$. %where
%$i=6, 9, 12, 14, 18, 21, 24, 36, 144, 168, 216, 342, 504$.

{\tiny
\[
G(C_{38}^6)=\left[\begin{array}{c}
    10000000000000000001010110011101101001\\
    01000000000000000001010110011101010110\\
    00100000000000000000010101101000011000\\
    00010000000000000000010101100111100111\\
    00001000000001000001100111011001010100\\
    00000100000001000001100111100110101000\\
    00000010000001000001110000100101100001\\
    00000001000001000001110011011010101101\\
    00000000100001000000100101011001101110\\
    00000000010001000000100110011010100001\\
    00000000001001000001000110010000010010\\
    00000000000101000001000100111100001100\\
    00000000000011000000000010101111100001\\
    00000000000000100000110000010000110101\\
    00000000000000010000000011010011110110\\
    00000000000000001001100010001100101000\\
    00000000000000000101010001111111011011\\
    00000000000000000011110010011111100010\\
    00000000000000000000001101011111101101
\end{array}\right]
\]
\[
G(C_{38}^9)=\left[\begin{array}{c}
    10000000000000000000011101010111111101\\
    01000000000000000000011101010111000010\\
    00100000000000000001010011111101100010\\
    00010000000000000001010011110010010001\\
    00001000000000000001100001101110010111\\
    00000100000000000001100001010001100111\\
    00000010000000000000111101011000110110\\
    00000001000000000000111110100111111001\\
    00000000100000000001101000100100111001\\
    00000000010000000001101011100111111010\\
    00000000001000000000001011101101000101\\
    00000000000100000000000100011110110101\\
    00000000000010000001001111010010110110\\
    00000000000001000001000000100010110110\\
    00000000000000100001111011011010100010\\
    00000000000000010001001000011001101110\\
    00000000000000001000100100011001010010\\
    00000000000000000100010111101010101101\\
    00000000000000000011111111000000001100
\end{array}\right]
\]
\[
G(C_{38}^{12})=\left[\begin{array}{c}
    10000000000000000011000110001010110100\\
    01000000000000000011000110001010001011\\
    00100000000000000011010110100010000101\\
    00010000000000000011010110100001001010\\
    00001000000000000100000111000001101101\\
    00000100000000000100000111001101100010\\
    00000010000000000010010000100101101011\\
    00000001000000000010010011011010010111\\
    00000000100000000111000011111010101011\\
    00000000010000000111000100101100111010\\
    00000000001000000001000101001010011110\\
    00000000000100000001010001110011111010\\
    00000000000010000000000110011001011000\\
    00000000000001000011000110100101010100\\
    00000000000000100101000101010101010111\\
    00000000000000010110010010010011001111\\
    00000000000000001111010000011100000110\\
    00000000000000000000100000101100110110\\
    00000000000000000000001000101010101101
\end{array}\right]
\]
\[
G(C_{38}^{14})=\left[\begin{array}{c}
    10000000000000000000101001101100111000\\
    01000000000000000000101001101100000111\\
    00100000000000000000111010000100010100\\
    00010000000000000000111010000111100111\\
    00001000000000000000010111111000111111\\
    00000100000000000000010111110111001111\\
    00000010000000000000110101010011000000\\
    00000001000000000000110101100011111100\\
    00000000100000000001111110100110000110\\
    00000000010000000001111101010110110101\\
    00000000001000000000001010111110011011\\
    00000000000100000000000110000010011011\\
    00000000000010000001001110000010010100\\
    00000000000001000001000001001101100111\\
    00000000000000100001010010011000100010\\
    00000000000000010001100010010111101101\\
    00000000000000001000001110011011011110\\
    00000000000000000100111101100100100001\\
    00000000000000000011111100110000001111
\end{array}\right]
\]
\[
G(C_{38}^{18})=\left[\begin{array}{c}
    10000000000000000100000111100100111000\\
    01000000000000000100000111100100000111\\
    00100000000000000101000100010111001111\\
    00010000000000000101000100011000110011\\
    00001000000000000000100101101000100001\\
    00000100000000000000100101010111010001\\
    00000010000000000101100000010010111010\\
    00000001000000000101100011101101111001\\
    00000000100000000100110101101110001001\\
    00000000010000000100110110101101001010\\
    00000000001000000101010110100111111001\\
    00000000000100000101010111110010100010\\
    00000000000010000100010010011000000101\\
    00000000000001000100010011001110101101\\
    00000000000000100101100010010110100100\\
    00000000000000010101010001010101100100\\
    00000000000000001100110011110011110000\\
    00000000000000000011110001100110100100\\
    00000000000000000000001110100110101011
\end{array}\right]
\]
\[
G(C_{38}^{21})=\left[\begin{array}{c}
    10000000000000000100001010111101011100\\
    01000000000000000100001010111101100011\\
    00100000000000000100100101000100101101\\
    00010000000000000100100101001011011110\\
    00001000000000000100101001000010111001\\
    00000100000000000100101001111101001001\\
    00000010000000000000000011001101011011\\
    00000001000000000000000000110010011011\\
    00000000100000000000110111100010110101\\
    00000000010000000000110100100001111010\\
    00000000001000000000110101111000100111\\
    00000000000100000000111010001011010100\\
    00000000000010000000010000010100111001\\
    00000000000001000000011111100100111001\\
    00000000000000100100001101100011101110\\
    00000000000000010100111110100000100010\\
    00000000000000001100110011110011110000\\
    00000000000000000010011110010011101110\\
    00000000000000000001100001010011101101
\end{array}\right]
\]
\[
G(C_{38}^{24})=\left[\begin{array}{c}
    10000000000000000000110110010111100110\\
    01000000000000000000110110010111011001\\
    00100000000000000000101110101111010001\\
    00010000000000000000101110101100101101\\
    00001000000001000000011101100111001100\\
    00000100000001000000011101101000000000\\
    00000010000001000000010000010011110011\\
    00000001000001000000010000101111001111\\
    00000000100000000000000111111001011110\\
    00000000010000000000000100110110010001\\
    00000000001001000000010001110101011110\\
    00000000000101000000011101110110010010\\
    00000000000011000000001100110000111100\\
    00000000000000100000111000101011010111\\
    00000000000000010000001011010111100111\\
    00000000000000001000111101010001111001\\
    00000000000000000100001110101101111010\\
    00000000000000000010100101111001101101\\
    00000000000000000001011001001010101101
\end{array}\right]
\]
\[
G(C_{38}^{36})=\left[\begin{array}{c}
    10000000000000000001110100011110101001\\
    01000000000000000001110100011110010110\\
    00100000000000000001010110110111011011\\
    00010000000000000001010110111000011000\\
    00001000000000000001000101010001101101\\
    00000100000000000001000101101110100010\\
    00000010000000000000011100101101111010\\
    00000001000000000000011111010010000110\\
    00000000100000000001001001010001111010\\
    00000000010000000001001010010010110110\\
    00000000001000000000101010011000001010\\
    00000000000100000000100101101011110101\\
    00000000000010000001101110100111111001\\
    00000000000001000001100001010111000101\\
    00000000000000100001111110010000101011\\
    00000000000000010001001101010011010100\\
    00000000000000001000100001010011101011\\
    00000000000000000100010010100000010111\\
    00000000000000000011111111000000111100
\end{array}\right]
\]
\[
G(C_{38}^{144})=\left[\begin{array}{c}
    10000000000000000000110101000110111110\\
    01000000000000000000110101000110000001\\
    00100000000000000000010101101110110011\\
    00010000000000000000010101101101111100\\
    00001000000000000000010111111011001100\\
    00000100000000000000010111110111000011\\
    00000010000000000000110101010011110011\\
    00000001000000000000110101100000001100\\
    00000000100000000001101100100101101011\\
    00000000010000000001101111010101010100\\
    00000000001000000001110001100101101110\\
    00000000000100000001111101011001010010\\
    00000000000010000000110101011001101110\\
    00000000000001000000111010010101011101\\
    00000000000000100000111011000011000110\\
    00000000000000010000001011001100001010\\
    00000000000000001001100111000011001010\\
    00000000000000000101010100111100110101\\
    00000000000000000011111100110011111100
\end{array}\right]
\]
\[
G(C_{38}^{168})=\left[\begin{array}{c}
    10000000000000000001000010111010001000\\
    01000000000000000001000010111010110111\\
    00100000000000000101000100010111001111\\
    00010000000000000101000100011000001100\\
    00001000000000000001001100001100010100\\
    00000100000000000001001100110011100111\\
    00000010000000000100000111010000100111\\
    00000001000000000100000100101111011011\\
    00000000100000000101001100111011001010\\
    00000000010000000101001111111000001001\\
    00000000001000000100101111110010110101\\
    00000000000100000100100000000001110101\\
    00000000000010000101101011001101001001\\
    00000000000001000101100100111101110110\\
    00000000000000100101101100110000001111\\
    00000000000000010101000001100100100010\\
    00000000000000001100101101100100101110\\
    00000000000000000011100001010111011110\\
    00000000000000000000011110010111101101
\end{array}\right]
\]

\[
G(C_{38}^{216})=\left[\begin{array}{c}
    10000000000000000000001101101111100110\\
    01000000000000000000001101101111011001\\
    00100000000000000000111010000100101101\\
    00010000000000000000111010000111100010\\
    00001000000000000010011111111110011011\\
    00000100000000000010011111110010010100\\
    00000010000000000010111000100011001111\\
    00000001000000000010111011011100111100\\
    00000000100000000000001000111001010001\\
    00000000010000000000000111000101101101\\
    00000000001000000000111110110000110110\\
    00000000000100000000000010001100111010\\
    00000000000010000000101110011111110011\\
    00000000000001000010011101110000111010\\
    00000000000000100010001110010110100010\\
    00000000000000010000000010000110010111\\
    00000000000000001000010011110110100010\\
    00000000000000000110010000010101100100\\
    00000000000000000001110011010011111001
\end{array}\right]
\]

\[
G(C_{38}^{342})=\left[\begin{array}{c}
    10000000000001000000001010101111111011\\
    01000000000001000000001010101111000100\\
    00100000000000000100001101101001111010\\
    00010000000000000100001101100110111001\\
    00001000000001000100101001000000101011\\
    00000100000001000100101001111111100111\\
    00000010000001000000101011100010011110\\
    00000001000001000000101000011101101110\\
    00000000100001000000101000000110110011\\
    00000000010001000000101011000101111100\\
    00000000001001000000000010110001111010\\
    00000000000101000000001101000010000110\\
    00000000000011000000001111110000110000\\
    00000000000000100100100101001110000101\\
    00000000000000010100001001101011101101\\
    00000000000000001100101100010101100111\\
    00000000000000000010101001011000010001\\
    00000000000000000001001001111110110101\\
    00000000000000000000011111100110010111
\end{array}\right]
\]

\[
G(C_{38}^{504})=\left[\begin{array}{c}
    10000000000000010010011000100100010110\\
    01000000000000010010011000100100101001\\
    00100000000000010010010110111100011000\\
    00010000000000010010010110110011010100\\
    00001000000000000010011001010011110000\\
    00000100000000000010011001101100001100\\
    00000010000000010010000100000110000011\\
    00000001000000010010000111111010110000\\
    00000000100000000000010111100110101101\\
    00000000010000000000011000100101101110\\
    00000000001000010000010100011110100100\\
    00000000000100010000011111010101011101\\
    00000000000010010000010010001010111100\\
    00000000000001010010000000010101100010\\
    00000000000000110010011010100111100111\\
    00000000000000001000001100010110101110\\
    00000000000000000110011001111110111001\\
    00000000000000000001010010100011010010\\
    00000000000000000000110111001000000101
\end{array}\right]
\]
}

\end{document}